\pdfoutput=1
\documentclass[pdflatex,sn-nature,iicol]{sn-jnl}

\usepackage{graphicx}
\usepackage{multirow}
\usepackage{amsmath,amssymb,amsfonts}
\usepackage{amsthm}
\usepackage{mathrsfs}
\usepackage[title]{appendix}
\usepackage{xcolor}
\usepackage{textcomp}
\usepackage{manyfoot}
\usepackage{booktabs}
\usepackage{algorithm}
\usepackage{algpseudocode}
\usepackage{listings}
\usepackage{url}
\usepackage{array}

\hypersetup{hidelinks}
\graphicspath{{figures/}}

\lstdefinestyle{mystyle}{
    basicstyle=\ttfamily\small,
    breaklines=true,
    frame=single,
    columns=fullflexible,
    keepspaces=true
}
\title{QMClaw: A Scalable General-purpose Framework for Quantum Measurement and Control}

\author[1]{\fnm{Zhiqiang} \sur{Fan}}
\author[1]{\fnm{Haoran} \sur{He}}
\author[1]{\fnm{Ping} \sur{Lv}}
\author[1]{\fnm{Junchao} \sur{Wang}}
\author[2]{\fnm{Yaqiang} \sur{Sun}}
\author[1]{\fnm{Chenhui} \sur{Wang}}
\author[1]{\fnm{Hanshi} \sur{Zhao}}
\author[1]{\fnm{Geyuyan} \sur{Ma}}
\author[1]{\fnm{Haoran} \sur{Yang}}
\author[1]{\fnm{Pengyu} \sur{Han}}
\author[1]{\fnm{Xiangdong} \sur{Meng}}
\author[1]{\fnm{Lixin} \sur{Wang}}
\author[1]{\fnm{Feng} \sur{Yue}}
\author*[1]{\fnm{Weilong} \sur{Wang}}\email{wangwl19888@163.com}
\author*[1]{\fnm{Zheng} \sur{Shan}}\email{shanzhengzz@163.com}

\affil*[1]{\orgname{Laboratory for Advanced Computing and Intelligence Engineering}, \orgaddress{\city{Zhengzhou}, \country{China}}}
\affil[2]{\orgname{Institute of Automation, Chinese Academy of Sciences}, \orgaddress{\city{Beijing}, \country{China}}}

\abstract{
As quantum computing continues to scale, quantum measurement and control (QMC) are increasingly constrained by calibration workflow complexity and by requirements for low-latency execution, robust exception handling, and traceable workflow governance. Existing frameworks for QMC are specialized and task-specific, while language-model-based agents for QMC suffer from excessive latency and cannot satisfy the strict timing and control-density demands of large-scale quantum systems. Here we propose QMClaw, a general, workflow-oriented framework for QMC built, featuring a local-first, tool-governed, robust architecture. At its core is a RuleEngine-centered control layer that processes structured context, performs rule-based state transitions, and generates execution plans for typical calibration workflows. Language models are used only for natural-language interaction, high-level task understanding, and exception support, keeping the critical fast path efficient. We implement a single-qubit tune-up workflow as a demonstration and validation using real quantum device dataset. We also prove that the framework achieves quantitatively acceptable levels in terms of resource cost, LLM calling times and decision latency, enabling its practical deployment in large-scale quantum qubit measurement and control scenarios. This work presents a general workflow-oriented framework for QMC and provides evidence that rule-centered architectures are a promising design choice for scalable quantum-system calibration.
}

\keywords{quantum measurement and control, general-purpose framework, multi-functional agent system, LLM-based automation, state machines}
\begin{document}

\maketitle

\section{Introduction}

Quantum computing is entering the systems engineering stage~\cite{castelvecchi_ibm_2023, google_quantum_ai_and_collaborators_quantum_2024}, where quantum measurement and control acts as a vital bridge between quantum processors and application tasks~\cite{2026Near001,2025Robust002,2025TensorHyper003,2026Faithful004,2025Improved005,alexeev_quantum_2021}. In the noisy intermediate-scale quantum (NISQ) era~\cite{bharti_noisy_2022, preskill2018nisq,fu_heterogeneous_2021,liu_high_2025,tao_scalable_nodate,arute_quantum_2019}, quantum hardware suffers from decoherence, calibration drift, and device nonuniformity. Experimental procedures such as configuration, tuning, interpretation, and scheduling still rely heavily on expert manual operation. Existing script-based automation~\cite{fan2025automation, klimov_snake_2020, krantz_quantum_2019, schuff_fully_2024,xu_automatic_2023,qiao_unveiling_2025,marciniak_optimal_2022} accelerates isolated tasks but lacks a unified architecture for task management, runtime monitoring, closed loop adaptation, and traceable execution. Most existing quantum measurement and control frameworks are specialized, task specific, and tightly coupled to particular experimental setups, making them difficult to generalize across diverse hardware environments and workflow scenarios~\cite{2025Practical006, lennon_efficiently_2019}. A general, cross platform framework is urgently needed to coordinate multi stage calibration under real world experimental constraints.

Practical deployment of quantum systems~\cite{wack_quality_2021,xu_qubic_2021, zhang_m2cs_2024,wu_general_2016,fan_optimization_2025, ding_experimental_2024, he_control_2022} now depends more on robust experiment organization under stringent hardware limitations than on algorithm design. This is especially prominent in quantum measurement and control, which demands physically executable procedures, hardware aware parameterization, reliable backend interaction, and robust handling of noisy data. To this end, it is difficult to support adaptive closed loop control with rigid scripts. 

Recent advances in artificial intelligence~\cite{genois_quantum_2025,strikis_learning-based_2021,abolhasani_rise_2023,li2026llmqubit,stein2019automated,rapp_self-driving_2024,bennett_autonomous_2024, liao_machine_2024,alexeev_artificial_2025,ma_machine_2025} have opened new avenues for automating workflows in quantum measurement and control (QMC). Large language models (LLM)~\cite{boiko_autonomous_2023}, vision–language models (VLM), and agent-based systems\cite{reuer_realizing_2023} have shown considerable promise in interpreting goals, planning procedures, invoking instruments, and analyzing experimental outcomes. Such data-driven and semantic approaches offer clear advantages over rigid script-based pipelines, enabling more adaptive and intelligent operation of quantum platforms. Li et al. developed HAL~\cite{li2026llmqubit}, an LLM-based autonomous framework for superconducting qubit experiments that automates resonator characterization and literature reproduction. However, it depends heavily on domain knowledge and lacks cross-platform generality. Cao et al. proposed QCalEval~\cite{cao2026qcaleval}, the first benchmark for evaluating vision-language models (VLM) on quantum calibration plots, revealing that state-of-the-art VLM perform well in visual perception but lack domain knowledge for reliable diagnostic reasoning. They also released NVIDIA Ising Calibration 1, a domain-tuned VLM that improves zero-shot performance but still fails to close the multimodal in-context learning gap. Building on agent-based automation, Cao et al. propose the k-agents framework~\cite{cao2024agents}, an LLM driven multi-agent system that autonomously plans, executes, and analyzes quantum experiments via structured laboratory knowledge and agent based state machines, achieving expert level performance in superconducting qubit calibration.

Despite these benefits, existing LLM and VLM methods remain limited by latency, poor generalization across hardware setups, and unreliable interpretation of calibration diagnostics. In particular, unconstrained LLM reasoning introduces excessive overhead that violates the real-time constraints of large-scale quantum systems, while VLM often fail to robustly map calibration plots to actionable physical feedback. Furthermore, nearly all available frameworks are customized for specific experimental types or laboratory environments and lack a unified, general, and scalable architecture applicable across diverse quantum platforms and calibration tasks.

To address these gaps, we present QMClaw, a general-purpose, scalable, workflow-oriented framework for QMC centered on explicit rule-based orchestration rather than unrestricted model inference. QMClaw focuses on cross-platform scalability, deterministic fast-path execution, and traceable workflow management. It integrates lightweight LLM interaction for high-level task understanding and exception handling with a dedicated RuleEngine that maintains low-latency, context-aware control over calibration logic. This separation preserves real-time performance while retaining the flexibility of natural-language interfaces.

We instantiate QMClaw using a formal state-machine workflow for single-qubit calibration, a ubiquitous and critical calibration procedure in superconducting quantum systems. Using a complete calibration dataset, we validate the framework’s ability to sustain end-to-end workflow continuity, parameter context management, and closed-loop orchestration. We further conduct a systematic, evidence-tiered scalability study to characterize latency, concurrency, pipeline decomposition, coverage, and LLM invocation pressure under large-scale operating assumptions.
The contributions of this work are fourfold:

\begin{enumerate}
\item A general and scalable workflow-oriented framework (QMClaw) unifying natural-language interaction, structured tasking, tool governance, execution monitoring, and result reporting.
\item A RuleEngine-centered control architecture enabling lightweight, deterministic, traceable calibration workflow orchestration.
\item A fully validated state-machine implementation for single-qubit tune-up demonstrating robust closed-loop calibration implemented on real quantum hardware.
\item A systematic scalability analysis quantifying real-time performance, bottlenecks, and LLM efficiency under scale-up conditions.
\end{enumerate}

\section{Framework Design}
QMClaw is designed as a workflow-oriented execution framework for QMC. Rather than serving merely as a language-based assistant, it is intended to organize the full path from task intake to backend execution, result interpretation, and structured reporting. QMClaw preserves the advantages of local-first orchestration, modular tool chaining, and persistent state management, while extending the underlying system with multi-functional agent scheduling, retrieval-augmented generation (RAG)~\cite{lewis2020rag}, function calling and tool-use mechanisms inspired by recent language-agent research~\cite{schick2023toolformer,yao2023react}, and domain-specific workflow control. Concretely, all these functional capabilities are implemented as modular extensions built upon the framework’s native core orchestration engine: multi-functional agent scheduling expands the internal task dispatcher, RAG enhances the built-in knowledge retrieval module, function calling provides native interfaces for the underlying tool-execution layer, and the workflow control module adds a dedicated state-machine coordinator to complement the baseline workflow manager.
\begin{figure*}[ht]
    \centering
    \includegraphics[width=0.8\textwidth]{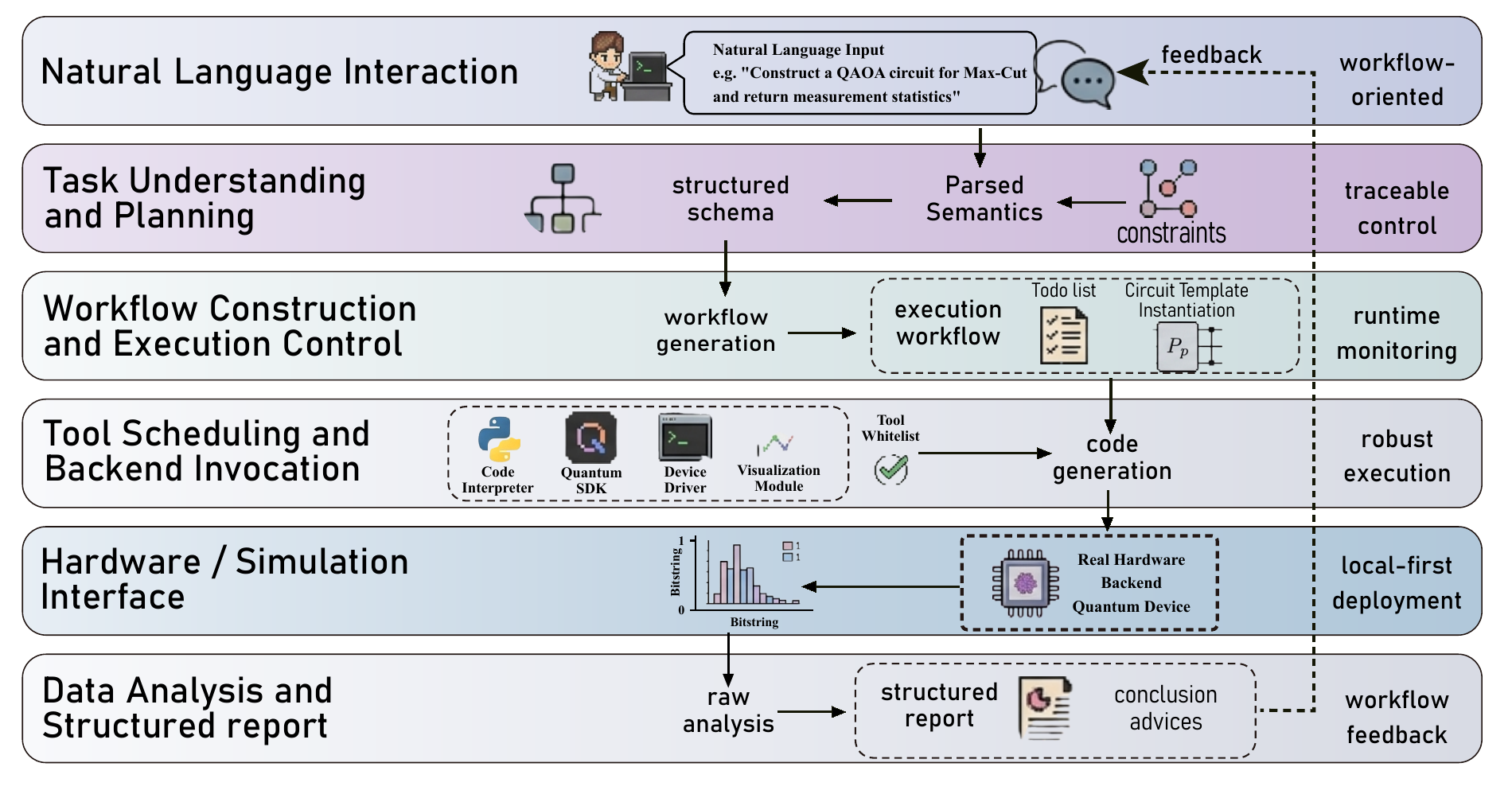}
    \caption{The overall design of QMClaw execution stack for scalable QMC}
    \label{fig:qmclaw_overall_execution_stack}
\end{figure*}

Such an architecture is motivated by several system-level requirements that distinguish QMC from ordinary conversational or coding assistants. First, the framework should support \emph{local-first deployment}, since measurement-and-control workflows often involve sensitive instrument commands, device states, calibration parameters, and experimental data that are better kept inside the laboratory environment. Second, it should provide \emph{heterogeneous tool orchestration}, because realistic quantum experiments span multiple software layers, including software development kits, simulators, hardware drivers, experiment-management scripts, and data-analysis utilities. Third, it should enable \emph{hardware-aware and constraint-aware execution}, so that generated procedures remain consistent with backend capabilities, parameter bounds, and device-specific restrictions. Finally, it should support \emph{traceable closed-loop control}, as quantum workflows are rarely one-shot procedures but instead involve repeated measurement, verification, fallback, and refinement. A usable intelligent system must therefore maintain explicit task states, monitor execution progress, and preserve intermediate results in a form suitable for auditing and recovery.

These requirements motivate an execution architecture in which explicit workflow state, bounded transition logic, and deterministic control remain on the fast path, while higher-level model reasoning is reserved for task interpretation and exceptional cases.

\subsection{Overall Architecture}
QMClaw is structured as a six-layer execution framework for QMC. Rather than presenting the system as a collection of isolated feature blocks, the architecture is designed to show how natural-language tasks are progressively transformed into structured workflows, executable actions, and interpretable outputs, all governed by explicit runtime control. Each layer of the framework plays a critical role in maintaining system flexibility, scalability, and seamless coordination. The six interconnected layers are as follows (See also Fig. \ref{fig:qmclaw_overall_execution_stack}):

\begin{enumerate}
    \item \textbf{Natural-Language Interaction Layer}: 
    This layer serves as the system’s interface to receive user requests in natural language. It performs initial intent recognition, leveraging natural language processing (NLP) to understand the user’s goals and objectives. The layer’s flexibility allows users to interact with the system in a way that abstracts away the underlying complexities of quantum control, enhancing the system's usability. This layer is scalable, accommodating a wide range of user inputs and queries.
    
    \item \textbf{Task Understanding and Planning Layer}: 
    Once the system understands the user's request, this layer extracts the necessary constraints and structures the goals into an actionable format. It maps tasks to predefined workflow templates and dynamic execution plans, accounting for dependencies and task hierarchies. This layer ensures that more complex, multi-step quantum experiments are broken down into manageable workflows. Its design allows the system to scale, supporting diverse tasks from simple calibration routines to complex multi-qubit operations, without needing to reconfigure the entire architecture.
    
    \item \textbf{Workflow Construction and Execution Control Layer}: 
    This layer takes the structured tasks from the planning layer and turns them into explicit workflow instances. It maintains runtime control logic, including state monitoring, dynamic task allocation, and adaptive decision-making. It ensures the robustness of the execution path, tracking progress and handling exceptions in real-time. This layer is fundamental to the system’s scalability, as it can orchestrate workflows across large quantum systems, adjusting dynamically as the number of qubits and complexity of tasks increase.
    
    \item \textbf{Tool Scheduling and Backend Invocation Layer}: 
    In this layer, the system schedules and invokes the required software tools, simulation engines, validation utilities, and execution interfaces. It allocates the necessary computational resources based on the specific task requirements, ensuring that the system efficiently uses its hardware and simulation resources. The system’s ability to scale is evident here, as it can manage large numbers of tool invocations concurrently, particularly in high-throughput quantum experiments.
    
    \item \textbf{Hardware/Simulator Interaction Layer}: 
    This layer serves as the bridge between the framework and the quantum hardware or simulators. It manages the communication with quantum backends, whether physical quantum devices or simulation environments, and ensures real-time data collection, feedback, and synchronization. This layer is crucial for scaling, as it must interface with varying hardware configurations, from small-scale devices to larger, more complex quantum systems, while ensuring that runtime metadata is consistently collected and monitored.
    
    \item \textbf{Data Analysis and Structured Reporting Layer}: 
    After executing the tasks, this layer interprets the outputs of the quantum experiments. It evaluates the validity of the results, analyzes performance metrics, and generates structured reports that include detailed execution traces, insights, and statistics. This layer supports the system's scalability by efficiently handling large volumes of experimental data and producing actionable insights for users. The reports generated can be used to fine-tune subsequent workflows, enabling continuous improvement of the system’s performance and ensuring transparency in decision-making.
\end{enumerate}

\begin{figure*}[ht]
    \centering
    \includegraphics[width=0.98\textwidth]{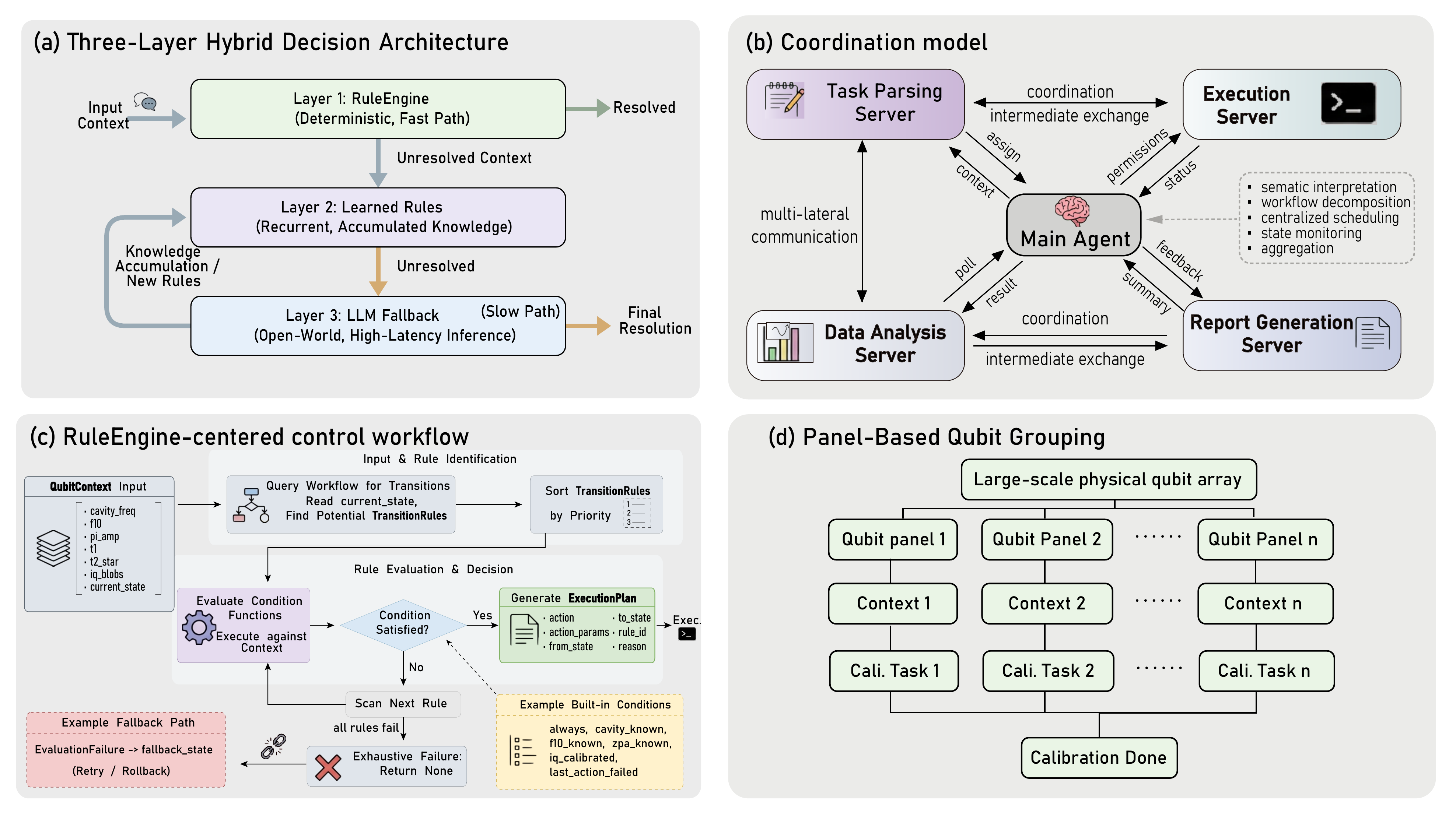}
    \caption{System design of QMClaw for scalable QMC. The integrated figure combines four coordinated views: (a) the three-layer hybrid decision architecture, (b) the centralized-scheduling and distributed-execution coordination model, (c) RuleEngine-centered control architecture in QMClaw, (d) Panel-based qubit grouping. Together, these views summarize how QMClaw organizes end-to-end execution, fast-path rule-based control, role-specialized coordination, and feedback-driven workflow progression.}
    \label{fig:qmclaw_system_design_overview}
\end{figure*}

This layered view emphasizes the end-to-end task flow, scalable workflow construction, backend execution, and feedback propagation across increasing task complexity. The multi-functional agent organization of QMClaw is discussed separately in the following subsection to maintain the focus of the system overview on the execution stack, while highlighting the coordination needed to scale the framework efficiently.

\subsection{Three-Layer Hybrid Decision Architecture}
A key architectural decision in QMClaw is that language models should not sit on the fast path of routine quantum measurement-and-control decisions. As system size grows, direct all-LLM control becomes difficult to justify from the perspectives of latency, determinism, cost, and auditability. QMClaw therefore adopts a three-layer hybrid decision architecture in which routine transitions remain governed by explicit rules, recurrent non-routine patterns are absorbed into learned state-machine rules, and language-model reasoning is reserved for rare unresolved cases outside the real-time critical loop.

The first layer, L1, is a minimal deterministic RuleEngine. It uses structured context and explicit transition rules to resolve standard calibration steps and generate bounded execution plans. L1 is intentionally small and inspectable: its role is not to memorize every possible experimental path, but to provide a stable fast path for canonical transitions and safety-relevant checks. Because L1 decisions are explicit, each action can be traced to a prior state, a rule condition, and a planned next state.

The second layer, L2, is a mined state-machine layer. It captures recurrent workflow structures observed in calibration records, including two-state loops, self-loops, and loop-exit transitions. Examples include frequency-scan and pulse-calibration loops such as IQRAW--PIPULSE, DRAG-related loops such as PIAMP--ALPHA, repeated evaluation states such as S21 or RAMSEY self-loops, and learned exit paths from these loops to downstream calibration steps. L2 is therefore the main mechanism by which QMClaw converts historical workflow regularities into reusable operational knowledge without requiring repeated model calls.

The third layer, L3, is a bounded LLM-assisted fallback. L3 is activated only when the structured context cannot be resolved by L1 or L2. Its role is exception support, diagnosis, explanation, and candidate rule acquisition rather than direct high-frequency control. When an L3-resolved case recurs and is validated, it can be materialized into L2 as a new learned transition pattern. This creates a knowledge-accumulation pathway: common behavior remains deterministic, recurrent exceptions become rules, and only the long tail remains model assisted.

At the architectural level, this hierarchy separates \emph{workflow governance} from \emph{open-ended interpretation}. Workflow governance is handled through structured context, explicit state transitions, tool-level execution boundaries, and auditable state updates. Open-ended interpretation is used only where it adds value, such as natural-language task understanding, result explanation, anomaly description, or rare fallback.

\subsection{Multi-functional Agent Scheduling Model}
QMClaw follows a centralized-scheduling, distributed-execution model. The main agent serves as the global coordination hub rather than directly carrying out all low-level operations. Its major responsibilities include receiving task requests, performing high-level semantic interpretation, decomposing tasks into workflow-compatible subtasks, assigning subtasks to specialized subagents, managing tool permissions, polling execution status, and aggregating final results.

Subagents are designed as modular execution units, each aligned with one or more functional responsibilities. Typical subagent roles include task-parsing agents, execution agents, data-analysis agents, and report-generation agents. Accordingly, the multi-functional agent design of QMClaw is role-specialized: the subagents are not treated as fully independent intelligent entities, but as functionally differentiated execution roles under a common orchestration mechanism. Such role specialization is particularly suitable for quantum measurement-and-control workflows, where parsing, execution, analysis, and reporting involve different constraints, tool interfaces, and output forms. Through this organization, the system distributes workload across specialized roles while preserving centralized oversight. Figure~\ref{fig:qmclaw_system_design_overview} (b) consolidates the main architectural views used in this paper, including the overall execution stack, the three-layer hybrid decision principle, the multi-functional agent coordination pattern, and the runtime mapping of a representative workflow.

Table~\ref{tab:subagents} summarizes representative subagent types in QMClaw.

\begin{table}[!t]
\caption{Representative Subagent Roles in QMClaw}
\label{tab:subagents}
\centering
\begin{tabular}{p{1.6cm}p{3.2cm}p{1.7cm}}
\toprule
\textbf{Subagent Type} & \textbf{Primary Responsibility} & \textbf{Typical Tooling} \\
\midrule
Task-parsing subagent & Intent recognition, task decomposition, case retrieval & RAG knowledge base \\
Execution subagent & Laboratory command execution and backend interaction & MCP server, \texttt{sq\_*} tools \\
Data-analysis subagent & Fitting, evaluation, and result interpretation & fitting utilities, XEB-related analysis \\
Report-generation subagent & Structured reporting and visualization & markdown/
 chart generation \\
\bottomrule
\end{tabular}
\end{table}

\subsection{Rule-Engine-Centered Workflow Design}

The central architectural question behind QMClaw is not simply how to automate an individual calibration script, but how a quantum measurement-and-control system should be organized when task scale, workflow complexity, and future qubit counts all increase. Under that viewpoint, the key problem is to maintain an explicit, low-latency, and auditable control layer that can drive routine workflow progression without requiring a language model to reason over the full experimental history at every step.

This paper addresses that problem through a RuleEngine-centered workflow design. The RuleEngine is responsible for reading the current runtime context, evaluating transition rules, and generating the next executable action. In this design, language-model-based components remain useful for task interpretation, knowledge retrieval, and exception-level assistance, but routine workflow control is moved onto an explicit rule-governed path. Single-qubit tune-up is used as a representative operational scenario because it contains multistage sequencing, evolving calibration parameters, measurement-contingent branching, and rollback requirements that are typical of broader measurement-and-control settings.

\subsection{Panel-Based Qubit Grouping}
To address the substantial LLM computational overhead incurred with the continuous expansion of qubit scale, this work adopts a panel partitioning grouping strategy for distributed qubit management. All physical qubits are divided into independent qubit panels according to hardware topology and functional correlation.

Each panel serves as a unified basic processing unit. All qubits within one panel are encapsulated into an integrated context batch and input into the large language model for centralized intelligent calibration decision-making. This grouping mode effectively compresses redundant context information, avoids scattered independent invocation for single qubits, and drastically cuts down total LLM invocation times, communication latency and resource consumption.

The constructed panel-level context specification acts as a unified interactive contract, which covers panel qubit quantity, hardware connection constraints, experimental shot configuration, calibration objectives, error tolerance criteria and result feedback rules.

Following the layered workflow, the system first splits large-scale global calibration demands into multiple subtasks corresponding to individual panels. Each panel task is matched with executable calibration templates and execution schemes. After completing in-panel measurement and control operations, raw data and operational metadata are returned uniformly. The analysis module then evaluates calibration validity of the whole panel following preset rules such as signal-to-noise ratio thresholds and parameter convergence bounds.

This framework also embeds closed-loop adaptive adjustment logic oriented to panel units. Calibration execution exceptions, hardware constraint conflicts and unsatisfactory in-panel calibration accuracy will trigger intra-panel strategy revision and parameter optimization. A limited retry mechanism is configured to ensure bounded iteration efficiency. By concentrating contextual information within panels rather than delivering massive full-scale qubit data to LLM at once, the proposed grouping architecture significantly improves context utilization efficiency, lowers overall inference cost, and strongly enhances the scalability of intelligent calibration systems toward large-scale quantum hardware.

\section{Validation}

\begin{figure*}[t]
    \centering
    \includegraphics[width=0.7\textwidth]{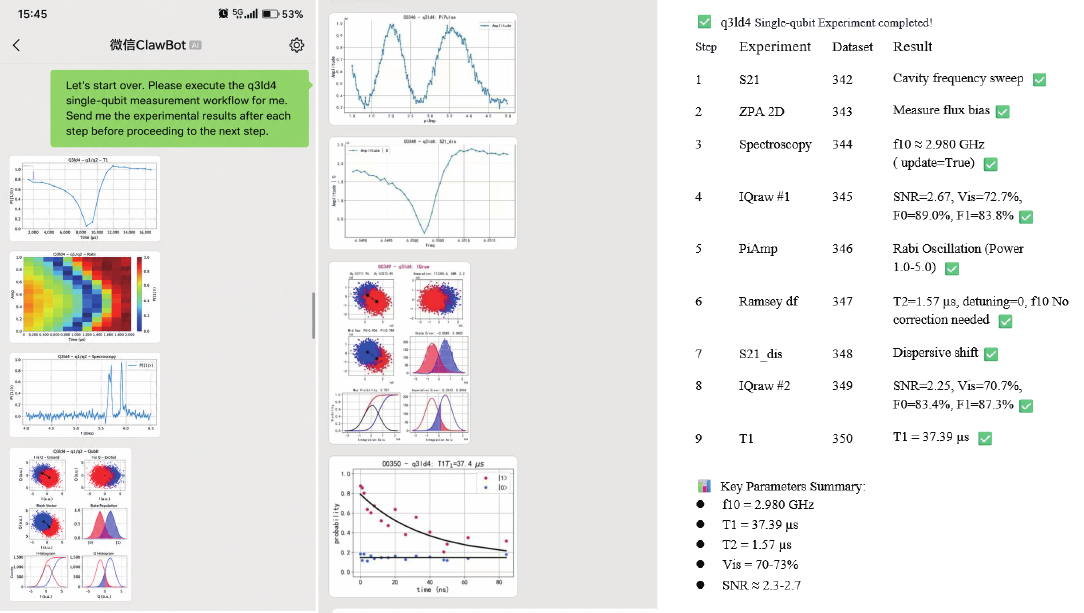}
    \caption{Automated quantum measurement and control workflow executed by the QMClaw intelligent agent. The experimental instructions were sent to the QMClaw agent through the WeChat official account interface, including resonator frequency scanning, flux bias optimization, qubit spectroscopy, readout calibration, Rabi oscillation, Ramsey coherence detection, and relaxation time $T_1$ measurement. The agent autonomously completed all experimental tasks and returned real-time measurement results, graphical data, and key performance metrics of the superconducting qubit.}
    \label{fig:natural_language_validation}
\end{figure*}
\subsection{Natural Language Control Validation}
We implemented single quantum measurement and control tasks via natural language control, which demonstrates that the proposed framework possesses sufficient intelligence to independently complete task execution and corresponding data analysis (see Fig. \ref{fig:natural_language_validation} for detail).

To validate this capability in a practical scenario, we connected the QMClaw agent to a WeChat-based interface, enabling users to issue high-level, human-readable instructions such as ``Let's start over. Please execute the q3ld4 single-qubit measurement workflow for me. Send me the experimental results after each step before proceeding to the next step.'' The agent successfully parsed the natural language intent, decomposed the request into a step-by-step workflow, and invoked the corresponding quantum measurement and control primitives in sequence. As shown in the figure, the workflow included resonator frequency scanning, flux bias optimization, qubit spectroscopy, readout characterization, $\pi$-pulse calibration via Rabi oscillation, Ramsey interferometry for $T_2^*$ measurement, dispersive shift characterization, and $T_1$ relaxation time measurement.

Throughout the process, the agent autonomously executed each experiment, collected raw data, performed on-the-fly analysis (such as fitting decay curves and calculating readout visibility and SNR), and returned both intermediate plots and key performance metrics. The complete sequence was finished without manual intervention, yielding a calibrated superconducting qubit with well-defined parameters ($f_{10} = 2.980$ GHz, $T_1 = 37.39$ $\mu$s, $T_2^* = 1.57$ $\mu$s, readout visibility $\approx 70$–$73\%$). This result verifies that the framework can bridge natural language commands and low-level quantum hardware operations, supporting closed-loop, end-to-end experiment orchestration.

\subsection{Historical Data Streaming Prediction Verification}
To validate the actual performance of the proposed framework, we conducted verification based on historical measurements and control experimental records of real quantum device. The dataset employed is session\_20251104, which contains 73 qubits and 2476 experimental steps, fully representing the actual operating sequence and sensor state variations in daily experiments. Under the established multi-layer Markov prediction strategy, the prediction accuracy is quantitatively evaluated on real experimental data.
The verification adopts the leave-one-qubit-out cross-validation protocol to ensure generalization capability. The prediction task is formulated as: given a historical sequence of experiment types and corresponding sensor quality states (IQ, PA, S21, T1 labeled as good/fair/poor/unknown), predict the next experiment type. The multi-layer strategy prioritizes second-order Markov exact matching, followed by first-order fallback, experiment-type-only fallback, and global fallback, which effectively balances prediction precision and robustness.
\begin{figure}[t]
    \centering
    \includegraphics[width=0.45\textwidth]{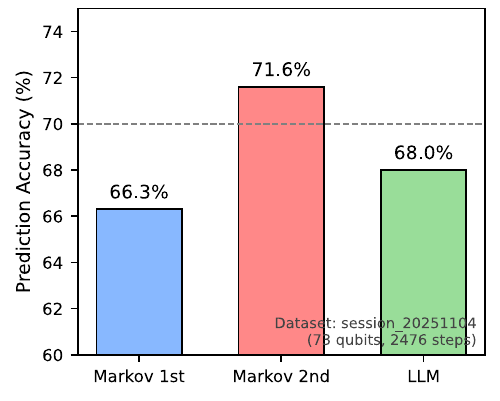}
    \caption{Historical streaming prediction accuracy comparison on real experimental data. The second-order Markov model achieves the highest accuracy of 71.6\% under the multi-layer prediction strategy, outperforming the first-order Markov model (66.3\%) and the LLM-based predictor (68\%). The leave-one-qubit-out cross-validation ensures generalization, with a small overfitting gap of 2.2\% observed. Dataset: session\_20251104 (73 qubits, 2476 steps).}
    \label{fig:streaming_pred_acc}
\end{figure}

As shown in Fig. \ref{fig:streaming_pred_acc}, experimental results demonstrate that the second-order Markov model achieves an overall prediction accuracy of 71.6\% on real streaming data, outperforming both the first-order Markov model (66.3\%) and the LLM-based sequence predictor (68\%). The overfitting gap between full training and cross-validation is only 2.2\%, confirming reliable generalization performance. The verification confirms that the proposed framework can accurately capture the real transition patterns of experimental operations, providing solid support for online streaming prediction applications.

\subsection{LLM Overhead Performance Verification}

We conducted statistical analysis on the number of LLM invocations and invocation latency of the framework, verifying its good compatibility for large-scale system expansion.

In the proposed three-layer hybrid architecture, the LLM is only employed as a fallback mechanism for rare and unforeseen states, rather than residing on the critical control path. 

\begin{figure}[t]
  \centering
  \includegraphics[width=0.48\textwidth]{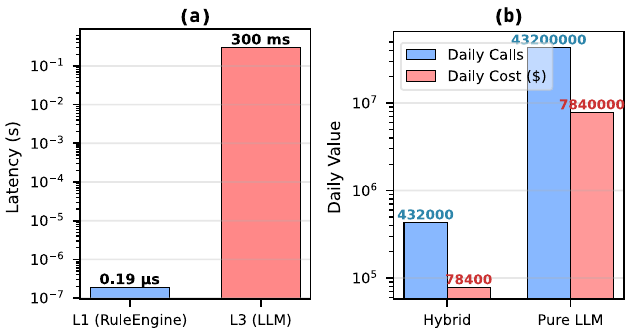}
  \caption{LLM Overhead Performance Verification. (a) Latency breakdown between Layer-1 RuleEngine (0.19\,$\mu$s) and LLM fallback (300\,ms), showing a 500,000$\times$ speed gap. (b) Daily LLM invocations and operational cost at 1M qubits: the hybrid architecture reduces calls by 100$\times$ and cost by 1000$\times$ compared to a pure-LLM baseline. Results confirm minimal LLM overhead and strong scalability for large-scale quantum systems.}
  \label{fig:llm_overhead}
\end{figure}

As verified in Fig.~\ref{fig:llm_overhead} (a), latency measurements show that the RuleEngine achieves an average latency of 0.19\,$\mu$s per evaluation, approximately 500,000 times faster than typical LLM inference (100--500\,ms per call). The overall system latency is dominated by the fast rule-based path, with only a tiny fraction of decisions subject to LLM-level delays. Under a million-qubit scenario, as shown in Fig.~\ref{fig:llm_overhead} (b), the daily number of LLM invocations is reduced from 43.2 million (pure-LLM baseline) to 432,000 (hybrid architecture), and further declines to 0.2\% of all decisions as rules accumulate over time. Cost analysis reveals a 1,000-fold reduction in daily operational cost compared with a pure-LLM design.

These results collectively verify that LLM overhead is well bounded and does not become a bottleneck. The framework maintains low latency, minimal invocation frequency, and stable cost growth under large-scale qubit control, demonstrating excellent scalability for million-qubit quantum systems.

\section{Discussion}
Experimental results verify that QMClaw serves as a systematic architecture for autonomous QMC rather than a simple benchmark platform. Practical quantum calibration workflows exhibit hybrid characteristics: most routine steps follow fixed rules, irregular procedures can be summarized from historical experimental data, and only a few extreme scenarios require LLM auxiliary decision-making. Correspondingly, QMClaw adopts a layered design, which arranges frequent operations into high-speed rule-based paths, integrates common abnormal cases into optimized state machine rules, and limits large language models to intent understanding, result interpretation and rare exception handling.

This design philosophy is distinct from pure LLM-driven quantum experiment agents. A critical limitation of relying solely on LLMs lies in their substantial invocation latency, which is incompatible with the low-latency requirements of real-time quantum calibration. Although prompt optimization can enhance sequence prediction capability, the inherent latency of LLM calls prevents them from supporting the rapid state transitions required in QMC workflows. Hence, LLMs are more suitable to act as auxiliary functional modules instead of core workflow controllers, while deterministic rules and empirical patterns—with negligible latency—dominate practical state transitions.

System-level latency tests further demonstrate the scalability of rule-based frameworks. The RuleEngine brings negligible time overhead, and the main pipeline delay derives from data interaction and task scheduling. This indicates that the core optimization direction for large-scale quantum experiment scheduling lies in data flow optimization and reducing redundant LLM invocations, rather than improving rule expression ability.

This work still has several limitations. Current workflow verification is mainly conducted on limited experimental sessions and single-type quantum devices, lacking sufficient cross-device and long-term stability tests. Besides, LLM-based prediction experiments require more samples for statistical verification, and panel aggregation strategy remains to be verified in actual distributed deployment. Moreover, this study mainly focuses on autonomous workflow scheduling, without comprehensive performance comparison with mainstream quantum control toolchains including Qiskit Experiments, C3 and Qibolab~\cite{kanazawa2023qiskit,wittler2021integrated,efthymiou2024qibolab}.

Aiming at the above shortcomings, multiple promising research directions are clarified. Firstly, expand the applicable scenarios of QMClaw to multi-qubit and coupled quantum device calibration, and optimize task scheduling and crosstalk suppression strategies. Secondly, realize online real-time rule updating, verification and rollback mechanisms. Thirdly, introduce reinforcement learning quantum control methods as constrained task modules~\cite{zhang2019when,nguyen2021deep,li2025robust,olle2024simultaneous}. Fourthly, utilize diffusion-based quantum signal generation technology to construct massive calibration datasets for algorithm stress testing. Finally, integrate visual language models specialized in quantum graph analysis~\cite{cao2026qcaleval} to enhance data interpretation ability, while maintaining the stable rule-based control core of QMClaw.

In summary, this work draws a clear design principle for intelligent QMC agents: artificial intelligence modules should be embedded inside standardized and constrained workflow frameworks, rather than directly participating in core real-time control as unrestricted decision executors. QMClaw provides both a feasible system architecture and empirical evidence for building scalable intelligent quantum experiment platforms based on structured context management, deterministic state transition and bounded intelligent assistance.

\section{Conclusion}
The core design of QMClaw revolves around a six-layer execution stack that enables end-to-end transformation of natural-language user requests into executable workflows, backend operations, and interpretable reports, with each layer contributing to system flexibility, scalability, and coordination. Complementing this layered architecture are three key design innovations: a three-layer hybrid decision architecture that separates routine rule-governed control (L1 RuleEngine, L2 mined state-machine) from LLM-assisted fallback (L3), a centralized-scheduling distributed-execution multi-functional agent model, and a RuleEngine-centered workflow design paired with panel-based qubit grouping to mitigate LLM computational overhead.

Experimental validation confirms the effectiveness and scalability of QMClaw. Natural language control validation demonstrates the framework’s ability to independently execute QMC tasks and analyze results. Historical data streaming prediction verification shows that the multi-layer Markov strategy achieves a prediction accuracy of 71.6\%, outperforming first-order Markov models and LLM-based predictors while maintaining strong generalization. LLM overhead performance verification further proves that the hybrid architecture reduces LLM invocations by 100 times and operational cost by 1000 times compared to pure-LLM designs, with the RuleEngine delivering sub-microsecond latency—five hundred thousand times faster than LLM inference—ensuring the framework remains efficient even for million-qubit quantum systems.

In summary, QMClaw resolves the key challenges of scaling intelligent QMC systems, including latency, determinism, cost, and auditability, by balancing structured rule-based control with targeted LLM assistance. Its modular design and scalable mechanisms make it well-suited for evolving quantum hardware, from small-scale devices to large-scale quantum systems. Future work will focus on expanding the RuleEngine’s rule library through automated learning, optimizing panel-based grouping strategies for diverse hardware topologies, and integrating more advanced anomaly detection to further enhance the framework’s robustness and adaptability in complex quantum experimental environments.

\section*{Acknowledgment}
 The authors thank the laboratory team and collaborators who contributed to the underlying QMC environment and experimental data collection.

The authors would like to acknowledge the assistance of large language models during manuscript preparation and experimental prototyping. The Minimax M2.7 large language model was utilized to support experimental auxiliary analysis; the Gemini model was adopted for sample image generation; Doubao and ChatGPT were employed to polish and refine the linguistic expression of this paper. All AI-generated contents have been carefully reviewed, revised and verified by the authors, who take full responsibility for all viewpoints, technical descriptions and experimental conclusions presented in this work.
\section*{Funding Declaration}
This work is supported by National Key Research and Development Program of China (Grant No. 2023YFB4502500). 

\section*{Competing Interests}
The authors declare no competing interests. 

\bibliography{qmclaw_refs}

@article{cao2024agents,
  author = {Cao, S. and Zhang, Z. and Alghadeer, M. and others},
  title = {Agents for Self-Driving Laboratories Applied to Quantum Computing},
  journal = {arXiv preprint},
  eprint = {2412.07978},
  archivePrefix = {arXiv},
  year = {2024}
}

@misc{cao2026qcaleval,
      title={QCalEval: Benchmarking Vision-Language Models for Quantum Calibration Plot Understanding}, 
      author={Shuxiang Cao and Zijian Zhang and Abhishek Agarwal and Grace Bratrud and Niyaz R. Beysengulov and Daniel C. Cole and Alejandro Gómez Frieiro and Elena O. Glen and Hao Hsu and Gang Huang and Raymond Jow and Greshma Shaji and Tom Lubowe and Ligeng Zhu and Luis Mantilla Calderón and Nicola Pancotti and Joel Pendleton and Brandon Severin and Charles Etienne Staub and Sara Sussman and Antti Vepsäläinen and Neel Rajeshbhai Vora and Yilun Xu and Varinia Bernales and Daniel Bowring and Elica Kyoseva and Ivan Rungger and Giulia Semeghini and Sam Stanwyck and Timothy Costa and Alán Aspuru-Guzik and Krysta Svore},
      year={2026},
      eprint={2604.25884},
      archivePrefix={arXiv},
      primaryClass={quant-ph},
      url={https://arxiv.org/abs/2604.25884}, 
}

@article{fan2025automation,
  author = {Fan, Zhiqiang and Wang, Weilong and Yue, Feng and others},
  title = {Design and Implementation of an Automated Measurement and Control System for Superconducting Quantum Chips},
  journal = {Journal of Information Engineering University},
  volume = {26},
  number = {1},
  pages = {51--56},
  year = {2025},
  note = {in Chinese}
}

@article{lewis2020rag,
  author = {Lewis, Patrick and Perez, Ethan and Piktus, Aleksandra and others},
  title = {Retrieval-Augmented Generation for Knowledge-Intensive NLP Tasks},
  journal = {Advances in Neural Information Processing Systems},
  volume = {33},
  pages = {9459--9474},
  year = {2020}
}

@article{li2026llmqubit,
  author = {Li, S. and Miller, J. M. and Lee, P. J. and others},
  title = {Large Language Model-Assisted Superconducting Qubit Experiments},
  journal = {arXiv preprint},
  eprint = {2603.08801},
  archivePrefix = {arXiv},
  year = {2026}
}

@article{preskill2018nisq,
  author = {Preskill, John},
  title = {Quantum Computing in the NISQ Era and Beyond},
  journal = {Quantum},
  volume = {2},
  pages = {79},
  year = {2018},
  doi = {10.22331/q-2018-08-06-79}
}

@article{stein2019automated,
  author = {Stein, H. S. and Gregoire, J. M.},
  title = {Progress and Prospects for Accelerating Materials Science with Automated and Autonomous Workflows},
  journal = {Chemical Science},
  volume = {10},
  number = {42},
  pages = {9641--9649},
  year = {2019}
}

@article{zhang2019when,
  author  = {Zhang, Xiao-Ming and Wei, Zezhu and Asad, Raza and Yang, Xu-Chen and Wang, Xin},
  title   = {When does reinforcement learning stand out in quantum control? A comparative study on state preparation},
  journal = {npj Quantum Information},
  volume  = {5},
  pages   = {85},
  year    = {2019},
  doi     = {10.1038/s41534-019-0201-8},
  url     = {https://www.nature.com/articles/s41534-019-0201-8}
}

@article{nguyen2021deep,
  author  = {Nguyen, V. and Orbell, S. B. and Lennon, D. T. and Moon, H. and Vigneau, F. and Camenzind, L. C. and Yu, L. and Zumb{\"u}hl, D. M. and Briggs, G. A. D. and Osborne, M. A. and Sejdinovic, D. and Ares, N.},
  title   = {Deep reinforcement learning for efficient measurement of quantum devices},
  journal = {npj Quantum Information},
  volume  = {7},
  pages   = {100},
  year    = {2021},
  doi     = {10.1038/s41534-021-00434-x},
  url     = {https://www.nature.com/articles/s41534-021-00434-x}
}

@article{olle2024simultaneous,
  author  = {Olle, Jan and Zen, Remmy and Puviani, Matteo and Marquardt, Florian},
  title   = {Simultaneous discovery of quantum error correction codes and encoders with a noise-aware reinforcement learning agent},
  journal = {npj Quantum Information},
  volume  = {10},
  pages   = {126},
  year    = {2024},
  doi     = {10.1038/s41534-024-00920-y},
  url     = {https://www.nature.com/articles/s41534-024-00920-y}
}

@article{li2025robust,
  author  = {Li, Shengyong and Fan, Yidian and Li, Xiang and Ruan, Xinhui and Zhao, Qianchuan and Peng, Zhihui and Wu, Re-Bing and Zhang, Jing and Song, Pengtao},
  title   = {Robust quantum control using reinforcement learning from demonstration},
  journal = {npj Quantum Information},
  volume  = {11},
  pages   = {124},
  year    = {2025},
  doi     = {10.1038/s41534-025-01065-2},
  url     = {https://www.nature.com/articles/s41534-025-01065-2}
}

@article{kanazawa2023qiskit,
  author  = {Kanazawa, Naoki and Egger, Daniel J. and Ben-Haim, Yael and Zhang, Helena and Shanks, William E. and Aleksandrowicz, Gadi and Wood, Christopher J.},
  title   = {Qiskit Experiments: A Python package to characterize and calibrate quantum computers},
  journal = {Journal of Open Source Software},
  volume  = {8},
  number  = {84},
  pages   = {5329},
  year    = {2023},
  doi     = {10.21105/joss.05329},
  url     = {https://joss.theoj.org/papers/10.21105/joss.05329}
}

@article{wittler2021integrated,
  author  = {Wittler, Nicolas and Roy, Federico and Pack, Kevin and Werninghaus, Max and Roy, Anurag Saha and Egger, Daniel J. and Filipp, Stefan and Wilhelm, Frank K. and Machnes, Shai},
  title   = {Integrated Tool Set for Control, Calibration, and Characterization of Quantum Devices Applied to Superconducting Qubits},
  journal = {Physical Review Applied},
  volume  = {15},
  number  = {3},
  pages   = {034080},
  year    = {2021},
  doi     = {10.1103/PhysRevApplied.15.034080},
  url     = {https://link.aps.org/doi/10.1103/PhysRevApplied.15.034080}
}

@article{efthymiou2024qibolab,
  author  = {Efthymiou, Stavros and Orgaz-Fuertes, Alejandro and Carrazza, Stefano and others},
  title   = {Qibolab: an open-source hybrid quantum operating system},
  journal = {Quantum},
  volume  = {8},
  pages   = {1247},
  year    = {2024},
  doi     = {10.22331/q-2024-02-12-1247},
  url     = {https://quantum-journal.org/papers/q-2024-02-12-1247/}
}

@article{yao2023react,
  author    = {Yao, Shunyu and Zhao, Jeffrey and Yu, Dian and Du, Nan and Shafran, Izhak and Narasimhan, Karthik and Cao, Yuan},
  title     = {{ReAct}: Synergizing Reasoning and Acting in Language Models},
  journal   = {arXiv preprint},
  year      = {2023},
  url       = {https://openreview.net/forum?id=WE_vluYUL-X}
}

@article{schick2023toolformer,
  author    = {Schick, Timo and Dwivedi-Yu, Jane and Dess{\`i}, Roberto and Raileanu, Roberta and Lomeli, Maria and Zettlemoyer, Luke and Cancedda, Nicola and Scialom, Thomas},
  title     = {Toolformer: Language Models Can Teach Themselves to Use Tools},
  journal   = {Advances in Neural Information Processing Systems},
  volume    = {36},
  year      = {2023}
}

@article{2026Near001,
  title={Near-term fermionic simulation with subspace noise tailored quantum error mitigation},
  author={ Papi, Miha  and  Algaba, Manuel G.  and  Godinez-Ramirez, Emiliano  and Inés de Vega and  Auer, Adrian  and  Fedor imkovic, I. V.  and  Calzona, Alessio },
  journal={npj Quantum Information},
  volume={12},
  number={1},
  year={2026},
}

@article{2025Robust002,
  title={Robust spin-qubit control in a natural Si-MOS quantum dot using phase modulation},
  author={ Kuno, Takuma  and  Utsugi, Takeru  and  Ramsay, Andrew J.  and  Mertig, Normann  and  Lee, Noriyuki  and  Yanagi, Itaru  and  Mine, Toshiyuki  and  Kusuno, Nobuhiro  and  Mizokuchi, Raisei  and  Nakajima, Takashi },
  year={2025},
}

@article{2025TensorHyper003,
  title={TensorHyper-VQC: A Tensor-Train-Guided Hypernetwork for Robust and Scalable Variational Quantum Computing},
  author={ Qi, Jun  and  Yang, Chao Han Huck  and  Chen, Pin Yu  and  Hsieh, Min Hsiu },
  year={2025},
}

@article{2026Faithful004,
  title={Faithful and secure distributed quantum sensing under general-coherent attacks},
  author={ Bizzarri, Gabriele  and  Barbieri, Marco  and  Manrique, Mylenne  and  Parisi, Miranda  and  Bruni, Fabio  and  Gianani, Ilaria  and  Rosati, Matteo },
  journal={NPJ Quantum Information},
  volume={12},
  number={1},
  year={2026},
}

@article{2025Improved005,
  title={Improved Quantum Computation using Operator Backpropagation},
  author={ Fuller, Bryce  and  Tran, Minh C.  and  Lykov, Danylo  and  Johnson, Caleb  and  Rossmannek, Max  and  Wei, Ken Xuan  and  He, Andre  and  Kim, Youngseok  and  Vu, Dinh Duy  and  Sharma, Kunal },
  year={2025},
}

@article{2025Practical006,
  title={Practical techniques for high-precision measurements on near-term quantum hardware and applications in molecular energy estimation},
  author={ Korhonen, Keijo  and  Vappula, Hetta  and  Glos, Adam  and  Cattaneo, Marco  and Zimborás, Zoltán and  Borrelli, Elsi Mari  and  Rossi, Matteo A. C.  and García-Pérez, Guillermo and  Cavalcanti, Daniel },
  journal={NPJ Quantum Information},
  volume={11},
  number={1},
  year={2025},
}

@article{lennon_efficiently_2019,
	title = {Efficiently measuring a quantum device using machine learning},
	volume = {5},
	copyright = {2019 The Author(s)},
	issn = {2056-6387},
	url = {https://www.nature.com/articles/s41534-019-0193-4},
	doi = {10.1038/s41534-019-0193-4},
	language = {en},
	number = {1},
	urldate = {2025-05-21},
	journal = {npj Quantum Information},
	publisher = {Nature Publishing Group},
	author = {Lennon, D. T. and Moon, H. and Camenzind, L. C. and Yu, Liuqi and Zumbühl, D. M. and Briggs, G. A .D. and Osborne, M. A. and Laird, E. A. and Ares, N.},
	month = sep,
	year = {2019},
	note = {TLDR: This work presents measurements on a quantum dot device performed by a machine learning algorithm in real time, and shows that the algorithm outperforms standard grid scan techniques, reducing the number of measurements required and the measurement time by 3.7 times.},
	pages = {79},
}

@misc{klimov_snake_2020,
	title = {The {Snake} {Optimizer} for {Learning} {Quantum} {Processor} {Control} {Parameters}},
	url = {http://arxiv.org/abs/2006.04594},
	doi = {10.48550/arXiv.2006.04594},
	language = {en},
	urldate = {2025-08-22},
	publisher = {arXiv},
	author = {Klimov, Paul V. and Kelly, Julian and Martinis, John M. and Neven, Hartmut},
	month = jun,
	year = {2020},
	note = {arXiv:2006.04594 [quant-ph]},
}

@article{rapp_self-driving_2024,
	title = {Self-driving laboratories to autonomously navigate the protein fitness landscape},
	volume = {1},
	issn = {2948-1198},
	url = {https://www.nature.com/articles/s44286-023-00002-4},
	doi = {10.1038/s44286-023-00002-4},
	language = {en},
	number = {1},
	urldate = {2026-03-26},
	journal = {Nature Chemical Engineering},
	author = {Rapp, Jacob T. and Bremer, Bennett J. and Romero, Philip A.},
	month = jan,
	year = {2024},
	pages = {97--107},
}

@article{bennett_autonomous_2024,
	title = {Autonomous reaction pareto-front mapping with a self-driving catalysis laboratory},
	volume = {1},
	copyright = {2024 The Author(s), under exclusive licence to Springer Nature America, Inc.},
	issn = {2948-1198},
	url = {https://www.nature.com/articles/s44286-024-00033-5},
	doi = {10.1038/s44286-024-00033-5},
	language = {en},
	number = {3},
	urldate = {2026-03-26},
	journal = {Nature Chemical Engineering},
	publisher = {Nature Publishing Group},
	author = {Bennett, J. A. and Orouji, N. and Khan, M. and Sadeghi, S. and Rodgers, J. and Abolhasani, M.},
	month = mar,
	year = {2024},
	pages = {241--250},
}

@article{xu_qubic_2021,
	title = {{QubiC}: {An} {Open}-{Source} {FPGA}-{Based} {Control} and {Measurement} {System} for {Superconducting} {Quantum} {Information} {Processors}},
	volume = {2},
	copyright = {https://creativecommons.org/licenses/by/4.0/legalcode},
	issn = {2689-1808},
	shorttitle = {{QubiC}},
	url = {https://ieeexplore.ieee.org/document/9552516/},
	doi = {10.1109/TQE.2021.3116540},
	language = {en},
	urldate = {2025-08-22},
	journal = {IEEE Transactions on Quantum Engineering},
	author = {Xu, Yilun and Huang, Gang and Balewski, Jan and Naik, Ravi and Morvan, Alexis and Mitchell, Bradley and Nowrouzi, Kasra and Santiago, David I. and Siddiqi, Irfan},
	year = {2021},
	pages = {1--11},
}

@article{krantz_quantum_2019,
	title = {A quantum engineer's guide to superconducting qubits},
	volume = {6},
	copyright = {© 2019 Author(s).},
	issn = {1931-9401},
	url = {https://aip.scitation.org/doi/abs/10.1063/1.5089550},
	doi = {10.1063/1.5089550},
	language = {en},
	number = {2},
	urldate = {2023-02-20},
	journal = {Applied Physics Reviews},
	author = {Krantz, P. and Kjaergaard, M. and Yan, F. and Orlando, T. P. and Gustavsson, S. and Oliver, W. D.},
	month = jun,
	year = {2019},
}

@article{liao_machine_2024,
	title = {Machine learning for practical quantum error mitigation},
	volume = {6},
	copyright = {2024 The Author(s), under exclusive licence to Springer Nature Limited},
	issn = {2522-5839},
	url = {https://www.nature.com/articles/s42256-024-00927-2},
	doi = {10.1038/s42256-024-00927-2},
	language = {en},
	number = {12},
	urldate = {2025-06-18},
	journal = {Nature Machine Intelligence},
	publisher = {Nature Publishing Group},
	author = {Liao, Haoran and Wang, Derek S. and Sitdikov, Iskandar and Salcedo, Ciro and Seif, Alireza and Minev, Zlatko K.},
	month = dec,
	year = {2024},
}

@article{zhang_m2cs_2024,
	title = {{M2CS}: a microwave measurement and control system for large-scale superconducting quantum processors},
	volume = {33},
	issn = {1674-1056, 2058-3834},
	shorttitle = {M2cs},
	url = {https://dx.doi.org/10.1088/1674-1056/ad8a49},
	doi = {10.1088/1674-1056/ad8a49},
	language = {en},
	number = {12},
	urldate = {2025-06-05},
	journal = {Chinese Physics B},
	publisher = {Chinese Physical Society and IOP Publishing Ltd},
	author = {Zhang, Jiawei and Sun, Xuandong and Guo, Zechen and Yuan, Yuefeng and Zhang, Yubin and Chu, Ji and Huang, Wenhui and Liang, Yongqi and Qiu, Jiawei and Sun, Daxiong and Tao, Ziyu and Zhang, Jiajian and Guo, Weijie and Jiang, Ji and Linpeng, Xiayu and Liu, Yang and Ren, Wenhui and Niu, Jingjing and Zhong, Youpeng and Yu, Dapeng},
	month = dec,
	year = {2024},
}

@misc{schuff_fully_2024,
	title = {Fully autonomous tuning of a spin qubit},
	url = {http://arxiv.org/abs/2402.03931},
	doi = {10.48550/arXiv.2402.03931},
	language = {en},
	publisher = {arXiv},
	author = {Schuff, Jonas and Carballido, Miguel J. and Kotzagiannidis, Madeleine and Calvo, Juan Carlos and Caselli, Marco and Rawling, Jacob and Craig, David L. and van Straaten, Barnaby and Severin, Brandon and Fedele, Federico and Svab, Simon and Kwon, Pierre Chevalier and Eggli, Rafael S. and Patlatiuk, Taras and Korda, Nathan and Zumbühl, Dominik and Ares, Natalia},
	year = {2024},
	note = {arXiv:2402.03931 [cond-mat]},
}

@article{alexeev_artificial_2025,
	title = {Artificial intelligence for quantum computing},
	volume = {16},
	issn = {2041-1723},
	url = {https://www.nature.com/articles/s41467-025-65836-3},
	doi = {10.1038/s41467-025-65836-3},
	language = {en},
	number = {1},
	urldate = {2026-01-04},
	journal = {Nature Communications},
	author = {Alexeev, Yuri and Farag, Marwa H. and Patti, Taylor L. and Wolf, Mark E. and Ares, Natalia and Aspuru-Guzik, Alán and Benjamin, Simon C. and Cai, Zhenyu and Cao, Shuxiang and Chamberland, Christopher and Chandani, Zohim and Fedele, Federico and Hamamura, Ikko and Harrigan, Nicholas and Kim, Jin-Sung and Kyoseva, Elica and Lietz, Justin G. and Lubowe, Tom and McCaskey, Alexander and Melko, Roger G. and Nakaji, Kouhei and Peruzzo, Alberto and Rao, Pooja and Schmitt, Bruno and Stanwyck, Sam and Tubman, Norm M. and Wang, Hanrui and Costa, Timothy},
	month = dec,
	year = {2025},
}

@article{alexeev_quantum_2021,
	title = {Quantum computer systems for scientific discovery},
	volume = {2},
	issn = {2691-3399},
	url = {https://link.aps.org/doi/10.1103/PRXQuantum.2.017001},
	doi = {10.1103/PRXQuantum.2.017001},
	language = {en},
	number = {1},
	urldate = {2026-01-04},
	journal = {PRX Quantum},
	author = {Alexeev, Yuri and Bacon, Dave and Brown, Kenneth R. and Calderbank, Robert and Carr, Lincoln D. and Chong, Frederic T. and DeMarco, Brian and Englund, Dirk and Farhi, Edward and Fefferman, Bill and Gorshkov, Alexey V. and Houck, Andrew and Kim, Jungsang and Kimmel, Shelby and Lange, Michael and Lloyd, Seth and Lukin, Mikhail D. and Maslov, Dmitri and Maunz, Peter and Monroe, Christopher and Preskill, John and Roetteler, Martin and Savage, Martin J. and Thompson, Jeff},
	month = feb,
	year = {2021},
	pages = {17001},
}

@article{abolhasani_rise_2023,
	title = {The rise of self-driving labs in chemical and materials sciences},
	volume = {2},
	issn = {2731-0582},
	url = {https://www.nature.com/articles/s44160-022-00231-0},
	doi = {10.1038/s44160-022-00231-0},
	language = {en},
	number = {6},
	urldate = {2026-03-26},
	journal = {Nature Synthesis},
	author = {Abolhasani, Milad and Kumacheva, Eugenia},
	month = jan,
	year = {2023},
	pages = {483--492},
}

@article{boiko_autonomous_2023,
	title = {Autonomous chemical research with large language models},
	volume = {624},
	issn = {0028-0836, 1476-4687},
	url = {https://www.nature.com/articles/s41586-023-06792-0},
	doi = {10.1038/s41586-023-06792-0},
	language = {en},
	number = {7992},
	urldate = {2026-03-26},
	journal = {Nature},
	author = {Boiko, Daniil A. and MacKnight, Robert and Kline, Ben and Gomes, Gabe},
	month = dec,
	year = {2023},
	pages = {571--578},
}

@article{ma_machine_2025,
	title = {Machine Learning for Estimation and Control of Quantum Systems},
	volume = {12},
	copyright = {https://creativecommons.org/licenses/by/4.0/},
	issn = {2095-5138, 2053-714X},
	url = {https://academic.oup.com/nsr/article/doi/10.1093/nsr/nwaf269/8191249},
	doi = {10.1093/nsr/nwaf269},
	language = {en},
	number = {8},
	urldate = {2025-09-20},
	journal = {National Science Review},
	author = {Ma, Hailan and Qi, Bo and Petersen, Ian R and Wu, Re-Bing and Rabitz, Herschel and Dong, Daoyi},
	month = jul,
	year = {2025},
	pages = {nwaf269},
}

@article{reuer_realizing_2023,
	title = {Realizing a Deep Reinforcement Learning Agent for Real-Time Quantum Feedback},
	volume = {14},
	issn = {2041-1723},
	url = {https://www.nature.com/articles/s41467-023-42901-3},
	doi = {10.1038/s41467-023-42901-3},
	language = {en},
	number = {1},
	urldate = {2025-09-20},
	journal = {Nature Communications},
	author = {Reuer, Kevin and Landgraf, Jonas and Fösel, Thomas and O'Sullivan, James and Beltrán, Liberto and Akin, Abdulkadir and Norris, Graham J. and Remm, Ants and Kerschbaum, Michael and Besse, Jean-Claude and Marquardt, Florian and Wallraff, Andreas and Eichler, Christopher},
	month = nov,
	year = {2023},
}

@article{wu_general_2016,
	title = {General-Purpose Quantum Computers: Theory, Composition and Implementation},
	volume = {39},
	issn = {0254-4164},
	shorttitle = {General-Purpose Quantum Computers},
	url = {https://kns.cnki.net/KCMS/detail/detail.aspx?dbcode=CJFQ&dbname=CJFDLAST2017&filename=JSJX201612002},
	language = {en},
	number = {12},
	urldate = {2025-09-25},
	journal = {Chinese Journal of Computers},
	author = {Wu, Nan and Song, Fangmin and Li, Xiangdong},
	year = {2016},
	pages = {2429--2445},
}

@article{fu_heterogeneous_2021,
	title = {Quantum-Classical Heterogeneous Computing System for Noisy Intermediate-Scale Quantum Technology},
	volume = {58},
	issn = {1000-1239},
	url = {https://kns.cnki.net/KCMS/detail/detail.aspx?dbcode=CJFQ&dbname=CJFDLAST2021&filename=JFYZ202109006},
	language = {en},
	number = {9},
	urldate = {2025-09-25},
	journal = {Journal of Computer Research and Development},
	author = {Fu, Xiang and Zheng, Yuzhen and Su, Xing and Yu, Jintao and Xu, Weixia and Wu, Junjie},
	year = {2021},
	pages = {1875--1896},
}

@phdthesis{liu_high_2025,
	address = {Beijing},
	type = {Doctoral Dissertation},
	title = {High-Fidelity Manipulation and Scaling of Superconducting Transmon Qubits},
	language = {en},
	school = {Tsinghua University},
	author = {Liu, Pei},
	month = jun,
	year = {2025},
}

@phdthesis{tao_scalable_nodate,
	type = {Doctoral Dissertation},
	title = {Quantum Simulation Based on Scalable Superconducting Quantum Circuits},
	language = {en},
	school = {Southern University of Science and Technology},
	author = {Tao, Ziyu},
}

@article{fan_optimization_2025,
	title = {Efficiency Optimization of Superconducting Quantum Measurement and Control Based on Variable Trigger Mechanism},
	issn = {1001-9081},
	url = {https://kns.cnki.net/KCMS/detail/detail.aspx?dbcode=CAPJ&dbname=CAPJLAST&filename=JSJY20251114003},
	language = {en},
	urldate = {2026-01-14},
	journal = {Journal of Computer Applications},
	author = {Fan, Zhiqiang and Wang, Lixin and He, Haoran and Ma, Ge Yuyan and Yue, Feng},
	month = nov,
	year = {2025},
}

@article{xu_automatic_2023,
	title = {Automatic {Qubit} {Characterization} and {Gate} {Optimization} with \textit{{QubiC}}},
	volume = {4},
	issn = {2643-6809, 2643-6817},
	url = {https://dl.acm.org/doi/10.1145/3529397},
	doi = {10.1145/3529397},
	language = {en},
	number = {1},
	urldate = {2025-08-22},
	journal = {ACM Transactions on Quantum Computing},
	author = {Xu, Yilun and Huang, Gang and Balewski, Jan and Morvan, Alexis and Nowrouzi, Kasra and Santiago, David I. and Naik, Ravi K. and Mitchell, Brad and Siddiqi, Irfan},
	month = mar,
	year = {2023},
	note = {TLDR: This work develops a concise and automatic calibration protocol to characterize qubits and optimize gates using QubiC, which is an open source FPGA (field-programmable gate array)-based control and measurement system for superconducting quantum information processors.},
	pages = {1--12},
}

@article{ding_experimental_2024,
	title = {Experimental advances with the {QICK} ({Quantum} {Instrumentation} {Control} {Kit}) for superconducting quantum hardware},
	volume = {6},
	issn = {2643-1564},
	url = {https://link.aps.org/doi/10.1103/PhysRevResearch.6.013305},
	doi = {10.1103/PhysRevResearch.6.013305},
	language = {en},
	number = {1},
	urldate = {2025-08-22},
	journal = {Physical Review Research},
	author = {Ding, Chunyang and Di Federico, Martin and Hatridge, Michael and Houck, Andrew and Leger, Sebastien and Martinez, Jeronimo and Miao, Connie and I, David Schuster and Stefanazzi, Leandro and Stoughton, Chris and Sussman, Sara and Treptow, Ken and Uemura, Sho and Wilcer, Neal and Zhang, Helin and Zhou, Chao and Cancelo, Gustavo},
	month = mar,
	year = {2024},
	pages = {013305},
}

@article{he_control_2022,
	title = {Control {System} of {Superconducting} {Quantum} {Computers}},
	volume = {35},
	issn = {1557-1939, 1557-1947},
	url = {https://link.springer.com/10.1007/s10948-021-06104-5},
	doi = {10.1007/s10948-021-06104-5},
	language = {en},
	number = {1},
	urldate = {2025-08-22},
	journal = {Journal of Superconductivity and Novel Magnetism},
	author = {He, Yongcheng and Liu, Jianshe and Zhao, Changhao and Huang, Rutian and Dai, Genting and Chen, Wei},
	month = jan,
	year = {2022},
	pages = {11--31},
}

@article{castelvecchi_ibm_2023,
	title = {{IBM} releases first-ever 1,000-qubit quantum chip},
	volume = {624},
	url = {https://ideas.repec.org//a/nat/nature/v624y2023i7991d10.1038_d41586-023-03854-1.html},
	language = {en},
	number = {7991},
	urldate = {2025-08-22},
	journal = {Nature},
	publisher = {Nature},
	author = {Castelvecchi, Davide},
	year = {2023},
	pages = {238--238},
}

@article{bharti_noisy_2022,
	title = {Noisy intermediate-scale quantum ({NISQ}) algorithms},
	volume = {94},
	issn = {0034-6861, 1539-0756},
	url = {http://arxiv.org/abs/2101.08448},
	doi = {10.1103/RevModPhys.94.015004},
	language = {en},
	number = {1},
	urldate = {2022-03-14},
	journal = {Reviews of Modern Physics},
	author = {Bharti, Kishor and Cervera-Lierta, Alba and Kyaw, Thi Ha and Haug, Tobias and Alperin-Lea, Sumner and Anand, Abhinav and Degroote, Matthias and Heimonen, Hermanni and Kottmann, Jakob S. and Menke, Tim and Mok, Wai-Keong and Sim, Sukin and Kwek, Leong-Chuan and Aspuru-Guzik, Alán},
	month = feb,
	year = {2022},
}

@article{arute_quantum_2019,
	title = {Quantum supremacy using a programmable superconducting processor},
	volume = {574},
	issn = {0028-0836, 1476-4687},
	url = {http://www.nature.com/articles/s41586-019-1666-5},
	doi = {10.1038/s41586-019-1666-5},
	language = {en},
	number = {7779},
	urldate = {2022-03-14},
	journal = {Nature},
	author = {Arute, Frank and Arya, Kunal and Babbush, Ryan and Bacon, Dave and Bardin, Joseph C. and Barends, Rami and Biswas, Rupak and Boixo, Sergio and Brandao, Fernando G. S. L. and Buell, David A. and Burkett, Brian and Chen, Yu and Chen, Zijun and Chiaro, Ben and Collins, Roberto and Courtney, William and Dunsworth, Andrew and Farhi, Edward and Foxen, Brooks and Fowler, Austin and Gidney, Craig and Giustina, Marissa and Graff, Rob and Guerin, Keith and Habegger, Steve and Harrigan, Matthew P. and Hartmann, Michael J. and Ho, Alan and Hoffmann, Markus and Huang, Trent and Humble, Travis S. and Isakov, Sergei V. and Jeffrey, Evan and Jiang, Zhang and Kafri, Dvir and Kechedzhi, Kostyantyn and Kelly, Julian and Klimov, Paul V. and Knysh, Sergey and Korotkov, Alexander and Kostritsa, Fedor and Landhuis, David and Lindmark, Mike and Lucero, Erik and Lyakh, Dmitry and Mandrà, Salvatore and McClean, Jarrod R. and McEwen, Matthew and Megrant, Anthony and Mi, Xiao and Michielsen, Kristel and Mohseni, Masoud and Mutus, Josh and Naaman, Ofer and Neeley, Matthew and Neill, Charles and Niu, Murphy Yuezhen and Ostby, Eric and Petukhov, Andre and Platt, John C. and Quintana, Chris and Rieffel, Eleanor G. and Roushan, Pedram and Rubin, Nicholas C. and Sank, Daniel and Satzinger, Kevin J. and Smelyanskiy, Vadim and Sung, Kevin J. and Trevithick, Matthew D. and Vainsencher, Amit and Villalonga, Benjamin and White, Theodore and Yao, Z. Jamie and Yeh, Ping and Zalcman, Adam and Neven, Hartmut and Martinis, John M.},
	month = oct,
	year = {2019},
	pages = {505--510},
}

@article{google_quantum_ai_and_collaborators_quantum_2024,
	title = {Quantum error correction below the surface code threshold},
	volume = {638},
	copyright = {2024 The Author(s)},
	issn = {0028-0836, 1476-4687},
	url = {https://www.nature.com/articles/s41586-024-08449-y},
	doi = {10.1038/s41586-024-08449-y},
	language = {en},
	number = {8052},
	urldate = {2025-05-21},
	journal = {Nature},
	publisher = {Nature Publishing Group},
	month = aug,
	year = {2024},
	pages = {920--926},
}

@article{strikis_learning-based_2021,
	title = {Learning-based quantum error mitigation},
	volume = {2},
	issn = {2691-3399},
	url = {https://link.aps.org/doi/10.1103/PRXQuantum.2.040330},
	doi = {10.1103/PRXQuantum.2.040330},
	language = {en},
	number = {4},
	urldate = {2023-01-13},
	journal = {PRX Quantum},
	author = {Strikis, Armands and Qin, Dayue and Chen, Yanzhu and Benjamin, Simon C. and Li, Ying},
	month = nov,
	year = {2021},
	pages = {40330},
}

@article{genois_quantum_2025,
	title = {Quantum optimal control of superconducting qubits based on machine-learning characterization},
	volume = {24},
	doi = {10.1103/d9yg-d3qr},
	language = {en},
	number = {3},
	journal = {Physical Review Applied},
	author = {Genois, \'Elie},
	year = {2025},
}

@misc{wack_quality_2021,
	title = {Quality, speed, and scale: three key attributes to measure the performance of near-term quantum computers},
	shorttitle = {Quality, speed, and scale},
	url = {http://arxiv.org/abs/2110.14108},
	doi = {10.48550/arXiv.2110.14108},
	language = {en},
	urldate = {2026-01-04},
	publisher = {arXiv},
	author = {Wack, Andrew and Paik, Hanhee and Javadi-Abhari, Ali and Jurcevic, Petar and Faro, Ismael and Gambetta, Jay M. and Johnson, Blake R.},
	month = oct,
	year = {2021},
}

@article{qiao_unveiling_2025,
	title = {Unveiling the nature of graphs through quantum graphon learning},
	volume = {12},
	issn = {2056-6387},
	url = {https://www.nature.com/articles/s41534-025-01141-7},
	doi = {10.1038/s41534-025-01141-7},
	language = {en},
	number = {1},
	urldate = {2026-01-21},
	journal = {npj Quantum Information},
	author = {Qiao, Wenbo and Zhang, Peng and Zhao, Jiaming and Ran, Shi-Ju},
	month = nov,
	year = {2025},
	pages = {8--22},
}

@article{marciniak_optimal_2022,
	title = {Optimal metrology with programmable quantum sensors},
	volume = {603},
	issn = {0028-0836, 1476-4687},
	url = {https://www.nature.com/articles/s41586-022-04435-4},
	doi = {10.1038/s41586-022-04435-4},
	language = {en},
	number = {7902},
	urldate = {2026-03-04},
	journal = {Nature},
	author = {Marciniak, Christian D. and Feldker, Thomas and Pogorelov, Ivan and Kaubruegger, Raphael and Vasilyev, Denis V. and Van Bijnen, Rick and Schindler, Philipp and Zoller, Peter and Blatt, Rainer and Monz, Thomas},
	month = mar,
	year = {2022},
	pages = {604--609},
}

\end{document}